\documentclass{aa}
\usepackage{natbib,amssymb,times}

\usepackage{graphicx} 
\usepackage{color}
\def\gr{$\gamma$-ray}
\def\por{Porphyrion}
\def\aap{A\&A}

\def\apjl{ApJL}

\newcommand{\s}{\,{\rm s}}
\newcommand{\cm}{\,{\rm cm}}
\newcommand{\nG}{\,{\rm nG}}
\newcommand{\erg}{\,{\rm erg}}
\newcommand{\keV}{\,{\rm keV}}
\newcommand{\MeV}{\,{\rm MeV}}
\newcommand{\GeV}{\,{\rm GeV}}

\begin{document}

\title{Intergalactic magnetism in a $\gamma$-ray beam as a model of Porphyrion}

\author{Andrii Neronov$^{1,2}$, Franco Vazza$^{3,4}$, Axel Brandenburg$^{5,6,7,8}$, Chiara Caprini$^{9,10}$}

\institute{
Universit\'e Paris Cit\'e, CNRS, Astroparticule et Cosmologie, 75006 Paris, France
\and
Laboratory of Astrophysics, \'Ecole Polytechnique F\'ed\'erale de Lausanne, 1015 Lausanne, Switzerland
\and
Dipartimento di Fisica e Astronomia, Universita di Bologna, Via Gobetti 93/2, 40129 Bologna, Italy
\and
INAF Istituto di Radioastronomia, Via P. Gobetti 101, 40129 Bologna, Italy
\and
Nordita, KTH Royal Institute of Technology and Stockholm University, Hannes Alfv\'ens v\"ag 12, 10691 Stockholm, Sweden
\and
The Oskar Klein Centre, Department of Astronomy, Stockholm University, AlbaNova, 10691 Stockholm, Sweden
\and
McWilliams Center for Cosmology \& Department of Physics, Carnegie Mellon University, Pittsburgh, PA 15213, USA
\and
School of Natural Sciences and Medicine, Ilia State University, 3-5 Cholokashvili Avenue, 0194 Tbilisi, Georgia
\and
D\'epartement de Physique Th\'eorique and Center for Astroparticle Physics, Universit\'e de Gen\`eve, 
1211 Geneve 4, Switzerland
\and
Theoretical Physics Department, CERN, 1211 Geneva 23, Switzerland
}

\authorrunning{A. Neronov et al}
\titlerunning{$\gamma$-ray beam as a model of Porphyrion}

\date{\today, \hfill NORDITA-2024-041, CERN-TH-2024-187 }

\abstract
{
We estimate the magnetic field in the jets of the recently discovered 7 Mpc long \por\ system.
We use non-detection of the system in gamma-rays to derive a lower bound on the co-moving magnetic field strength at the level of $\sim 10\nG$.
This value is consistent with recent estimates of magnetic fields in the filaments of the Large Scale Structure.
We discuss the possibility that, instead of being the extreme case of a radio jet formation scenario, \por\  actually traces a very-high-energy \gr\ beam emitted by an active galactic nucleus.
In such a model, jets do not need to spread into the voids of the Large Scale Structure to appear straight on a very large distance range,
and several anomalies of the standard radio jet scenarios can be solved at once.
}
\keywords{}
\maketitle

\section{Introduction}

Outflows from Active Galactic Nuclei (AGN) may spread high-energy particles and magnetic fields over very large, Megaparsec-scale volumes in the intergalactic medium. The recent discovery of the 7~Mpc scale jet structure of the \por\ galaxy \citep{nature} at the redshift $z\simeq 0.9$ shows that the sizes of such outflows may reach a significant fraction of the size of the voids of the Large Scale Structure (LSS). If such jets penetrate into the voids, they may transport magnetic fields as well as relativistic particles into the voids.

Void magnetic fields may be measured using the technique of \gr\ astronomy through the effect of secondary \gr\ emission from electron-positron pairs deposited in the voids by very-high-energy \gr s \citep{2009PhRvD..80l3012N}. Observations with current generation \gr\ telescopes impose a lower bound on the void magnetic fields at the level of $10^{-17}$ -- $3\times 10^{-14}$~G \citep{MAGIC:2022piy,2010Sci...328...73N,2023ApJ...950L..16A}, depending on the assumptions on the duty cycle of the parent source of \gr s. The void magnetic fields may be of cosmological origin \citep{Durrer:2013pga}, and in this case the measurement of their properties would provide a useful cosmological probe. While the general consensus from simulations is that the collective activity of galaxies alone cannot magnetize voids to the level implied by the \gr\ lower bound \citep[e.g.][]{bond22,tj24}, the magnetic fields spread by radio galaxies like \por\ may
``pollute'' the voids and prevent measurement of the relic voids fields there.

The \por\ jets have a number of remarkable properties. They remain collimated over at least $\approx 3.5 \rm ~Mpc$ on both sides of the parent AGN. This is puzzling \citep{nature} because one generically expects development of  magneto-hydrodynamical (MHD) instabilities in the directed relativistic plasma flow, either intrinsic or induced by the prolonged entrainment of gas from the surrounding Inter-Galactic Medium (IGM). Straightness of the jet also implies a low velocity of the parent AGN and the absence of variations of the orientation of the axis of the supermassive black hole in its core over a prolonged period of its activity of $\sim 2 \rm ~Gyr$. Modelling of such a remarkably stable jet \citep{nature} suggests that the AGN should have been working at nearly the Eddington limit during this entire time. 

In what follows, we propose an alternative explanation for the formation of \por\ (also applicable to other similar objects) and we argue that the \por\ jets do not necessarily need to spread into the voids to keep their direction stable over a very large distance. We discuss the possibility that the observed structure is a jet-like feature produced {\it by a beam of very-high-energy \gr s producing electron-positron pairs in the IGM}. Such \gr\ beam traces do not need to transport charged particle plasma from the AGN nucleus nor do they need to sustain their own magnetic fields.

\section{Inverse Compton emission from \por}

The synchrotron radio emission from the \por\ jets is produced mostly by electrons with energies $E_e$ that may be inferred from the frequency of the synchrotron radiation,  
\begin{equation}
    \nu_s\simeq 100\left[\frac{B}{10\nG}\right]\left[\frac{E_e}{30\mbox{ GeV}}\right]^2\mbox{ MHz},
\end{equation}
if the magnetic field $B$ in the emission region is known.
For example, if the field is at the level of the fields found in the filaments of the LSS, $B\sim 10\nG$ \citep{Carretti:2022fqk}, the synchrotron-emitting electrons have energies in the 10--100~GeV range.

\begin{figure}
    \includegraphics[width=\columnwidth]{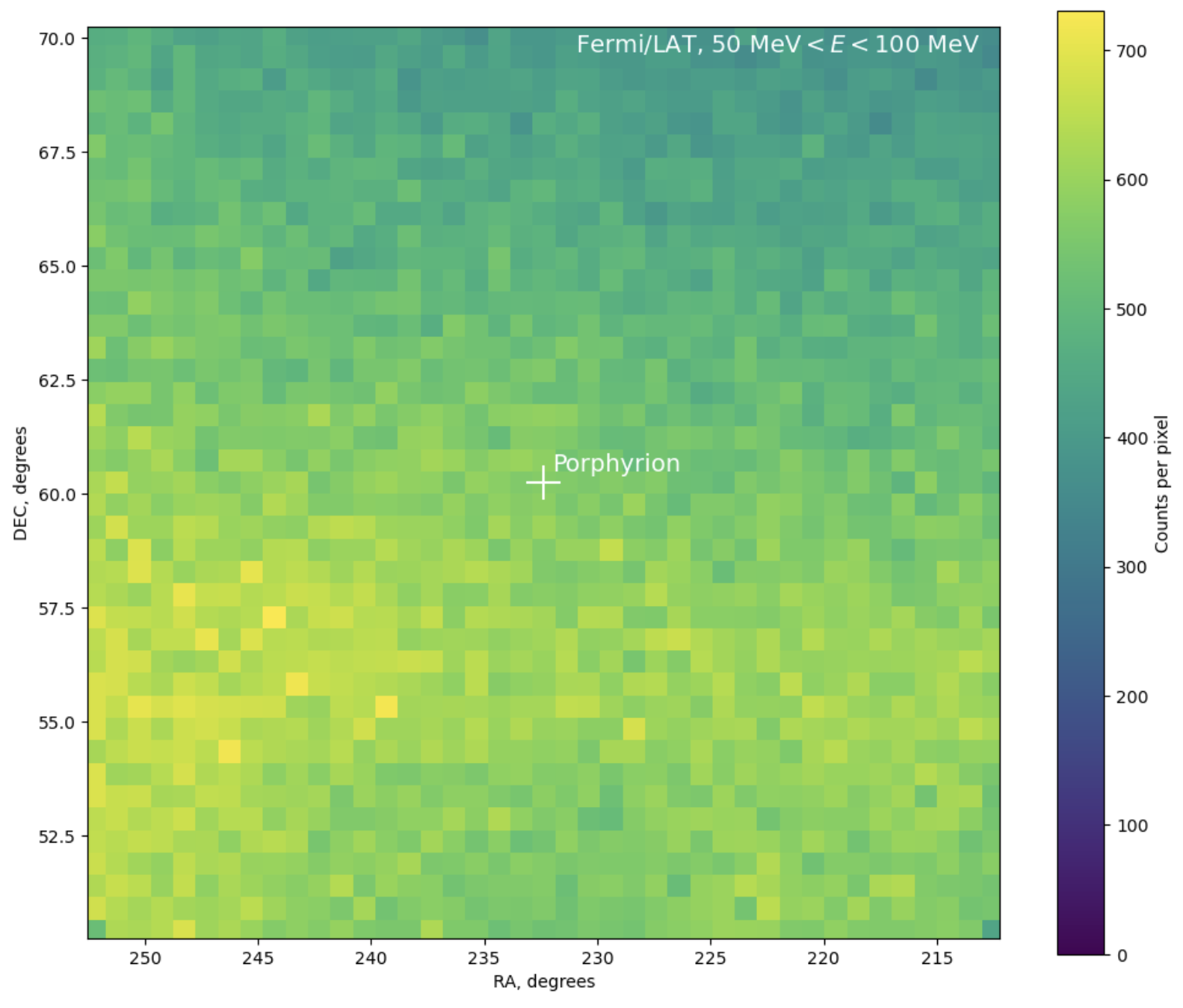}
    \includegraphics[width=\columnwidth]{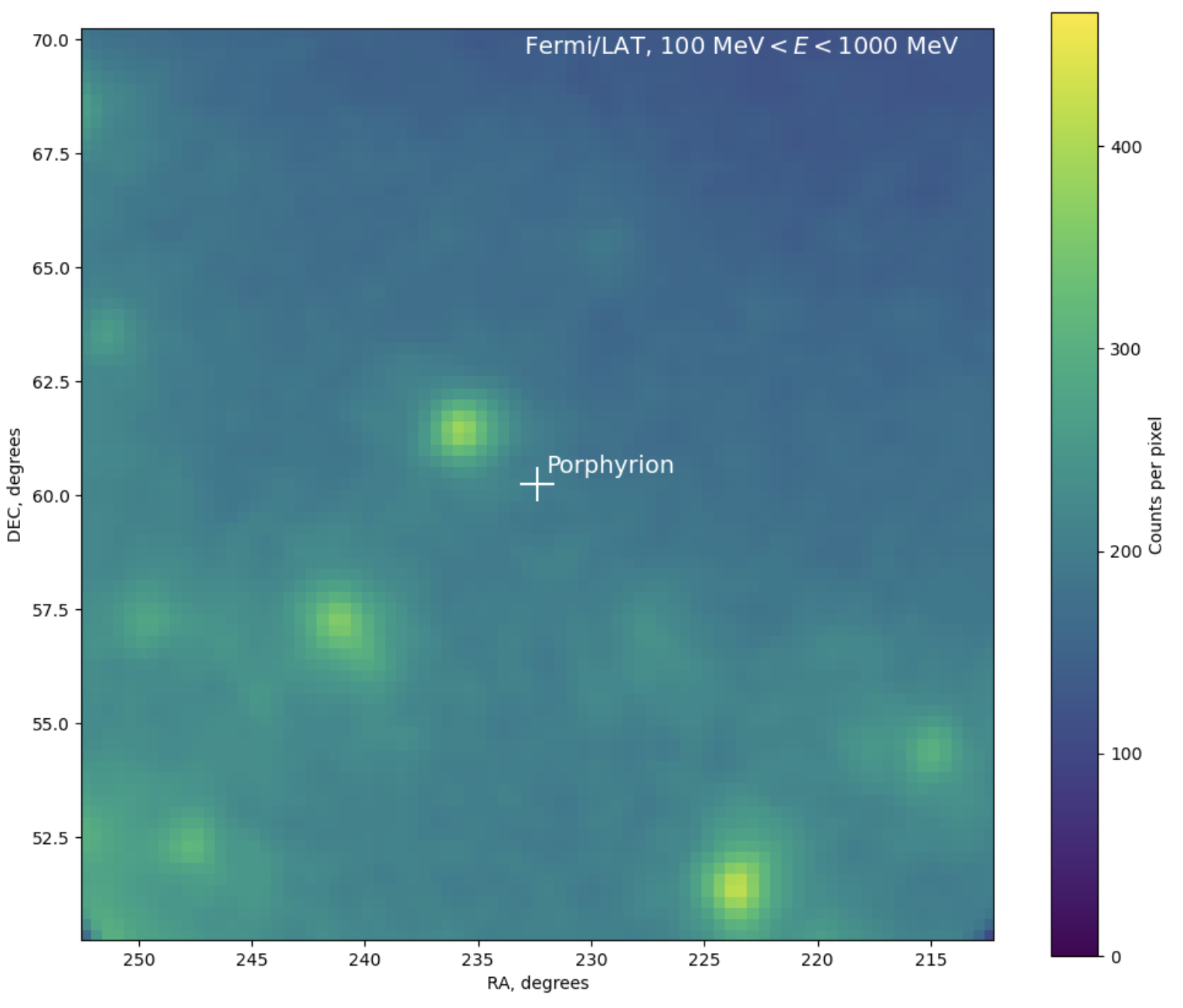}
    \caption{Fermi/LAT countmaps of the region around \por\ in the energy ranges $50\MeV<E_\gamma<100\MeV$ (top) and $100\MeV<E_\gamma<1000\MeV$ (bottom). The size of the \por\ source is smaller than the size of the cross at the source position. Brighter spots in the lower image are sources, other than \por, detected by Fermi/LAT. }
    \label{fig:lat_image}
\end{figure}

The same electrons also produce inverse Compton scattered photons from the CMB photons at the source redshift with energies \citep{2009PhRvD..80l3012N}
\begin{equation}
\label{IC}
    E_{\gamma}\simeq 6\left[\frac{E_{e}}{30\mbox{ GeV}}\right]^2\MeV.
\end{equation}

The ratio of the inverse Compton to synchrotron fluxes from the jet, $F_{IC}$ to $F_s$,
is determined by the ratio of the energy densities of the CMB and the magnetic field:
\begin{equation}
\label{eq:FICFs}
    \frac{F_{IC}}{F_s}=\frac{U_{CMB}}{U_B}\simeq 10^6\left[\frac{B}{10\nG}\right]^{-2}
\end{equation}
at redshift $z\simeq 0.9$,
where 
the energy density of the CMB 
is increased 
by a factor of $(1+z)^4\simeq 13$
compared to the local Universe.

The synchrotron flux of  \por\ is at the level $F_s\sim 10^{-16}\erg\cm^{-2}\s^{-1}$.  This suggests that the \gr\ flux may be as high as $F_{IC}\sim 10^{-10}\erg\cm^{-2}\s^{-1}$ in $E\sim 10\MeV$ band, yielding a very bright \gr\ source. Such a flux is potentially detectable with the Fermi Large Area Telescope (LAT) \citep{Atwood_2009} that has an energy threshold in the $E_\mathrm{thr,LAT}\sim 50\MeV$ range.

\section{Upper bound on the \gr\ flux of the jet}
\label{sec:upper_bound}

We use Fermi/LAT data publicly available through the Fermi Science Support 
Center\footnote{https://fermi.gsfc.nasa.gov/ssc/} to measure the flux of \por\ or to derive an upper limit on it. 
We consider \gr\ events with energies between $50\MeV$ and $100\GeV$ within $15^\circ$ distance from the source 
collected between 2008 and 2024.
We filter the events of P8R3\_SOURCEVETO\_V3 type using the {\tt gtselect-gtmktime} 
tool chain to retain only the highest quality events most likely expected to be \gr s. 

\begin{figure}
    \includegraphics[width=\columnwidth]{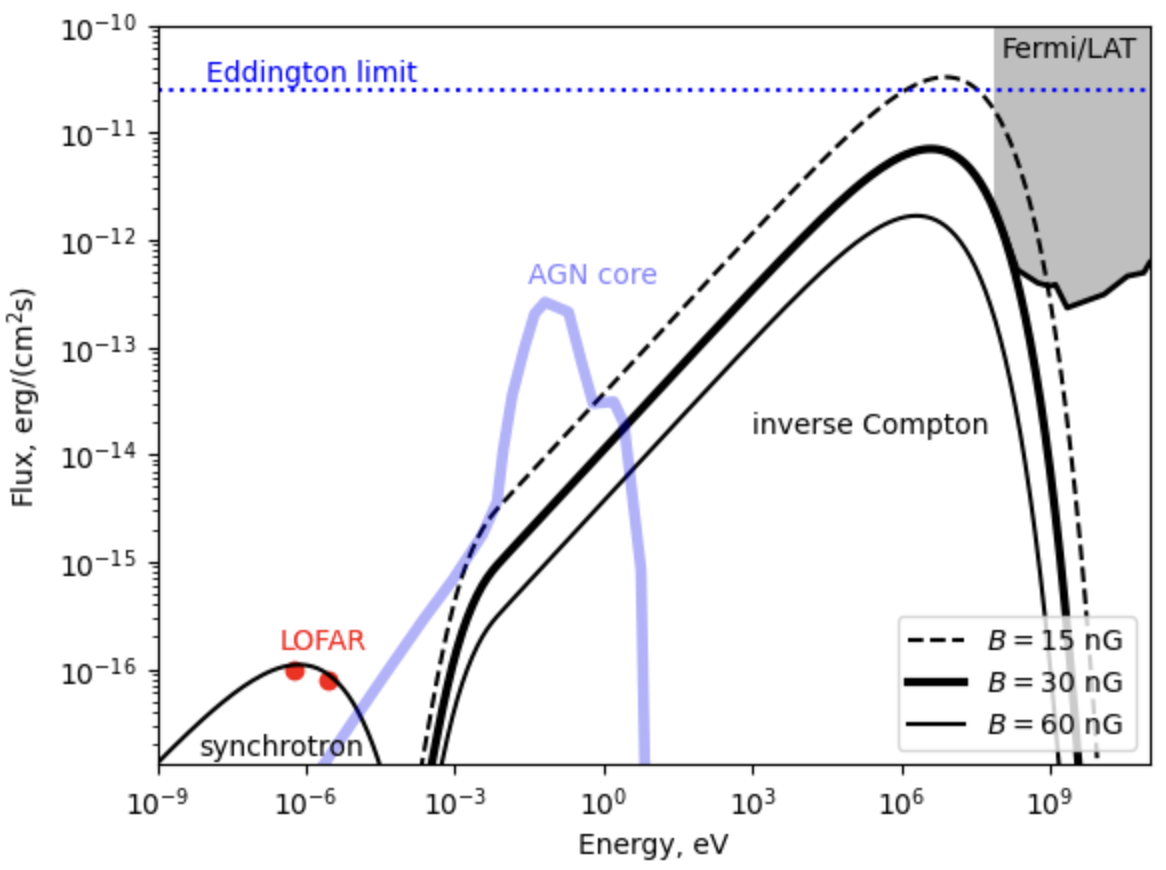}
    \caption{Spectral energy distribution of \por\ modelled as synchrotron and inverse Compton emissionlack data points show the radio flux of the jet. The grey-shaded region  shows an upper limit on \gr\ flux from Fermi/LAT. The light blue curve shows the host AGN spectral energy distribution model from \cite{nature}. Dashed horizontal line shows the level of flux expected from a source powered by a $10^9M_\odot$ black hole emitting at Eddington limit. Red, blue and green points show the expected inverse Compton flux level from  electrons emitting synchrotron radiation in the LOFAR frequency range, for different proper magnetic field levels.}
    \label{fig:sed}
\end{figure}

Figure~\ref{fig:lat_image} shows the countmaps of the \por\ region in Fermi/LAT in the energy ranges $50$--$100\MeV$ and $100\MeV$--$1\GeV$. As there is no excess in the direction of the source, we derive an upper limit on the source flux in logarithmically spaced energy bins, using the aperture photometry method. We extract the photon counts from circles of the radius $2^\circ$ ($1^\circ$) in the energy bins below (above) 1~GeV. We define the 95\% upper bound on the source flux as $2\sqrt{N}$, where $N$ is the number of counts in the source circle. Then, we estimate the exposure corrected for the fraction of the source flux contained within the fixed radius, using the {\tt gtexposure} tool. The resulting upper limit on the source flux is shown by the thick gray curve Fig. \ref{fig:sed}. 

\begin{figure*}
    \includegraphics[width=0.495\textwidth]{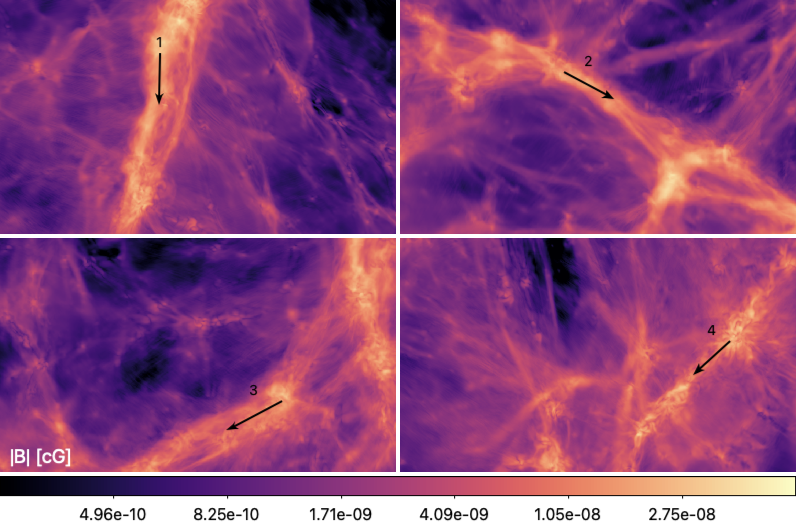}
       \includegraphics[width=0.495\textwidth]{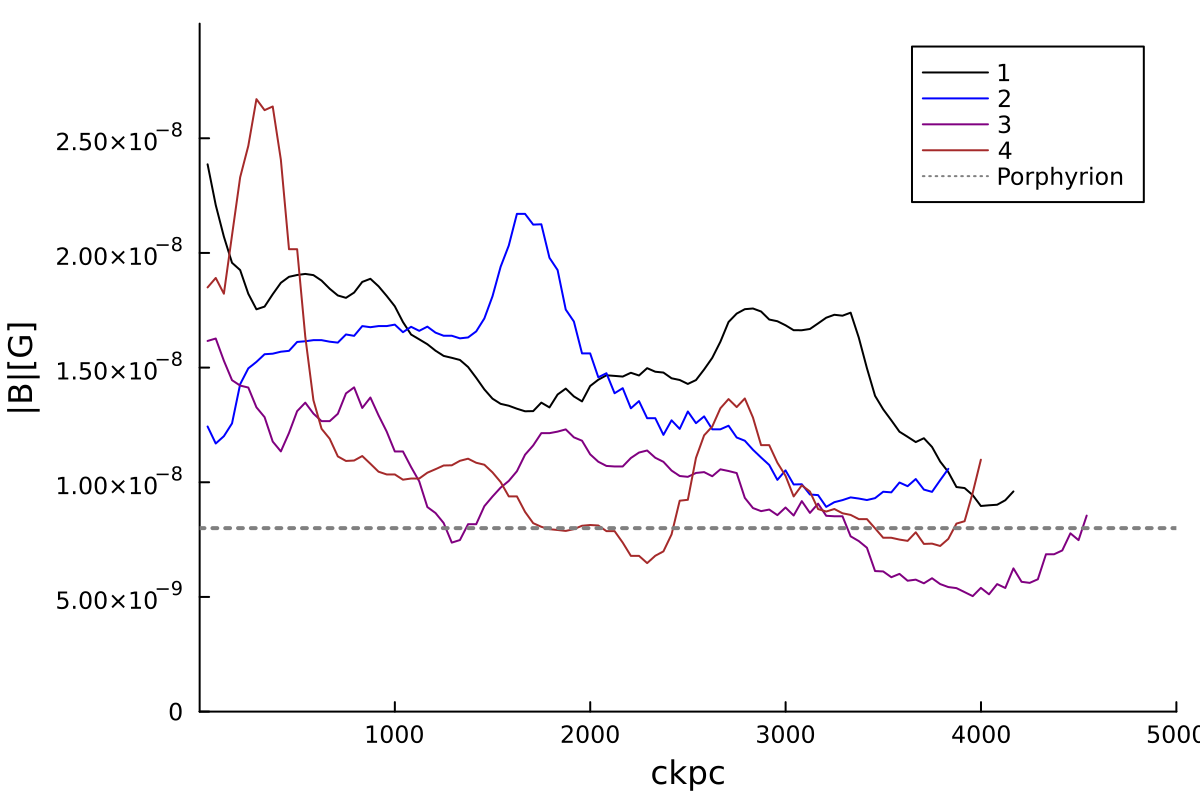}
    \caption{Left panels: mass-weighted mean conoving magentic field strength for four different filaments in a cosmological simulation at $z=1$. Right panels: corresponding profiles of comoving magnetic field strength through the same filaments, for a total extent compatible with the longest jet of Porphyrion.}
    \label{fig:Bsim}
\end{figure*}

To find an upper bound on the magnetic field, we model  synchrotron and inverse Compton emission from a cutoff powerlaw distribution of electrons,
\begin{equation}
    \frac{dN_e}{dE}\propto E^{-\Gamma}\exp\left(-\frac{E}{E_\mathrm{cut}}\right),\ \  E>E_\mathrm{min}.
\end{equation}
The seed photons for the inverse Compton scattering are provided by the CMB at the redshift of the source.
The resulting model is shown by the black solid line in Fig. \ref{fig:sed} for the minimal possible magnetic field strength,
\begin{equation}
    B_\mathrm{min}\simeq 30\nG.
\end{equation}
If the magnetic field were below this bound,
the inverse Compton flux would be observable in Fermi/LAT, because the inverse Compton component of the source spectrum (the right bump of the model curve)
would move in the upper-right direction and exceed the Fermi/LAT upper bound.  

\section{Comparison with magnetic fields in the filaments of the LSS}

The comoving minimal field strength,
\begin{equation}
    B_\mathrm{min,comoving}=\frac{B_\mathrm{min}}{(1+z)^2}\simeq 8\nG
\end{equation}
is close to the global estimate of the magnetic field in the filaments of the Large Scale Structure \citep[e.g.][]{Carretti:2022fqk,2024arXiv240616230M}. If the jets of \por\ are occasionally aligned along the filament direction, the field present in the synchrotron emission region may well be the background magnetic field in the LSS, rather than the field transported by the jet plasma. 

To more closely compare with the geometry of Porphyrion, we extracted from a large cosmological simulation at $z=1$ \citep[e.g.][]{va19} several 1-dimensional slices of magnetic fields starting from small-mass halos within approximately straight filaments, and out to a $\approx 3.5 \rm ~Mpc$  distance from the halo to reproduce the separation between the core and the outer lobes of Porphyrion.
 The simulation used a simple initially uniform ``primordial'' magnetic field seed of $0.1\nG$ (comoving) inserted at the start of the simulation ($z=50$) and evolved under ideal MHD conditions with dark and baryonic matter in the ENZO code. This simple model was shown to produce a very good match to the properties of stacked polarized radio emission from the cosmic web in the recent observations by \citet{2023SciA....9E7233V}. 
 
The general finding is that all paths have $B_{\rm comoving} \sim 10-15\nG$ magnetic fields for a large fraction of Porphyrion's projected extension, with frequent fluctuations of order $\times 2$ in magnetic field strength over scales of $\sim 10^2 \, \rm kpc$, corresponding to the typical size of substructures in filaments. 

This suggests that the jet may be contained within a filament of the Large Scale Structure.  The alignment of the jet with the filament direction need not to be strong. Indeed, the typical width of a filament is $r_f\sim 1.5$--$2$~Mpc \citep{filaments}, so that only a mild alignment, by an angle $\Theta\sim \arcsin(r_f/R_j)\simeq 30^\circ$ along the filament is needed for a jet of the length $R_j\simeq 3.5$~Mpc  to be contained in a filament.
Considering that longer jets require a better large scale alignment with surrounding filaments to produce detectable radio emission in this scenario, this might qualitatively explain why this phenomenon is relatively rare to observe. A quantitative exploration of the statistics of such configuration with full cosmological simulations is deferred to future work. 

\begin{figure*}
    \includegraphics[width=0.95\textwidth]{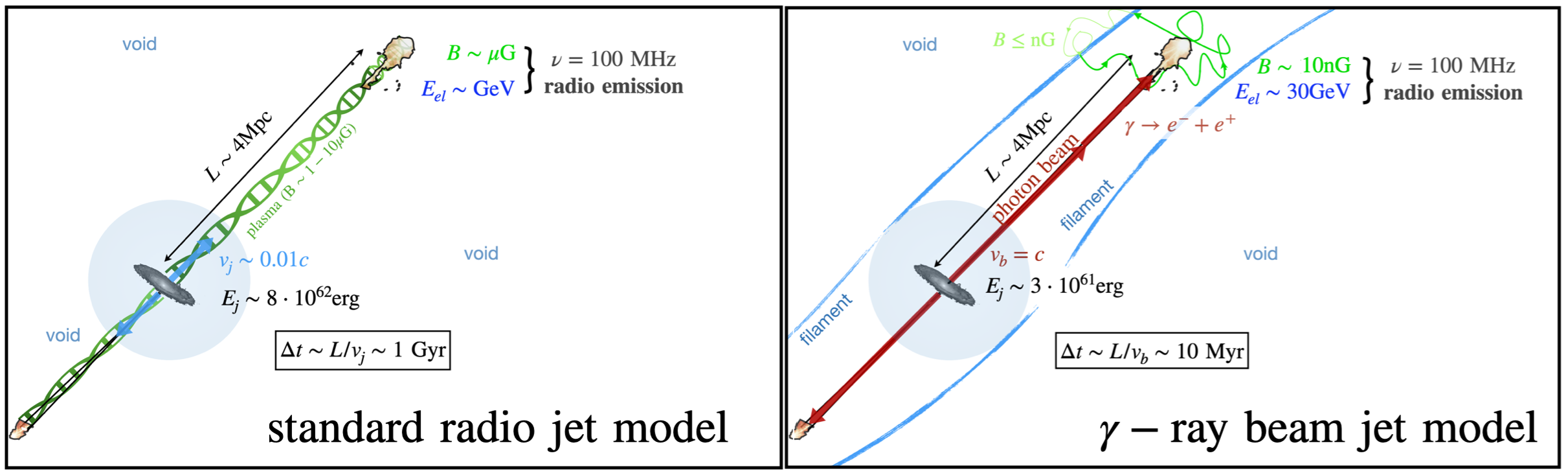}
    \caption{Cartoon views of the two physical models for \por\ discussed in this work.}
    \label{fig:sketch}
\end{figure*}

\section{Gamma-ray beam model of the system}

Observations with \gr\ telescopes have revealed a population ``TeV blazars'', AGNs that emit narrow beams of \gr s with energies in excess of 10~TeV \citep{Biteau:2020prb}. Alignment along the line of sight leads to strong Doppler boosting of \gr\ emission in blazars and facilitates their observations in the \gr\ band. In the absence of alignment, the \gr\ beam may not be detectable by \gr\ telescopes, but it should leave a trace in the intergalactic medium that is observable at lower energies \citep{Neronov_2002}. The trace is produced by synchrotron and inverse Compton emission by electrons and positrons deposited along the beam by the effect of pair production of \gr s interacting with the visible/infrared Extragalactic Background Light (EBL) \citep{PhysRev.155.1404}.  The mean free path of the  \gr s with energies $E_{\gamma_0}$ through the EBL at the redshift $z\sim 1$ \citep{Fitoussi_2017}
\begin{equation}
    D_{\gamma_0}\simeq \left\{
    \begin{array}{ll}
    10\mbox{ Mpc},& E_{\gamma_0}=10\mbox{ TeV},\\
    1\mbox{ Mpc},& E_{\gamma_0}=30\mbox{ TeV},
    \end{array}\right.
\end{equation}
is comparable to the length scale of \por-type beams
for $E_{\gamma_0}\gtrsim 10$~TeV. This means that all the energy of the \gr\ beam is converted into the energy of electrons and positrons deposited nearly homogeneously in the intergalactic medium on a several Mpc distance scale.

The electrons and positrons are initially injected in the energy range $E_e\sim (0.1$--$1) E_{\gamma_0}$. 
Once injected, these electrons and positron cool via inverse Compton scattering on Cosmic Microwave Background (CMB) at $z\simeq 1$ on the time scale
\begin{equation}
\label{eq:tcool}
    t_\mathrm{cool}\simeq 1.5\left[\frac{E_e}{30\mbox{ GeV}}\right]^{-1}\mbox{Myr}
\end{equation}
that becomes comparable to the light-crossing time of \por, which is $t_{lc}\simeq 10$~Myr for electrons with energies in the GeV energy range 
This leads to the formation of a low-energy tail of the electron-positron distribution.

If the source remains active at constant luminosity, the low energy tail spectrum is $dN/dE\propto E^{-2}$ \citep{Neronov_2002}. Otherwise, the shape of the electron spectrum depends on the history of the luminosity of the \gr\ beam on the cooling time scale. The spectrum of inverse Compton emission from an $E^{-2}$ distribution of electrons has the slope $dN_\gamma/dE\propto E^{-1.5}$ and the inverse Compton emission from TeV electrons is expected to be in the 10--100~GeV range.  Non-detection of this emission by Fermi/LAT suggests that the AGN core luminosity in very-high-energy \gr s should have been decreasing over the last millions of years so that only electrons and positrons that cooled down are surviving.
Otherwise, if the magnetic field in the jet is an order-of-magnitude  stronger than the lower bound derived above, the level of the inverse Compton flux would be below the Fermi/LAT upper bound, even if the source of the TeV \gr s would remain active till now.  In any case, the population of cooled electrons and positrons which remains along the trace of the \gr\ beam over $t\gtrsim 10$~Myr is expected to produce a radio synchrotron glow along the path of the \gr\ beam.  It is thus well possible that the straight jet of \por\ is a trace of a very-high-energy \gr\ beam emitted by the AGN core in the past ($\leq 10$--$15 ~\rm Myr$ ago). 

\section{Discussion}

The stability of jets in the relativistic regime has been thoroughly
studied with analytical and numerical approaches, to explore the role of jet parameters (bulk Lorentz factor, composition, density, temperature) and environment
\citep[e.g.][for recent reviews]{2019ARA&A..57..467B,2021NewAR..9201610K,2023JPlPh..89e9101P}.
A general finding is that the most important instabilities which can disrupt or de-collimate jets (like Kelvin-Helmholtz modes) grow faster in slower and hotter flows than in cold and fast ones. Faster jets with a high bulk Lorentz factor, $\Gamma$, develop short-wavelength and high-order modes that get amplified in the shear layer and eventually saturate the
growth of the instabilities, without loss of collimation, and generating a hot shear layer around the core of the jet \citep[][]{2022MNRAS.510.2084P}. Based on this, in order to explain the very high degree of stability of \por, one needs to assume a very relativistic jet ($\Gamma \sim 2$) travelling as a bullet, in an extremely low density environment.  It is, however, not clear if this picture, developed to interpret the jet stability on $\leq 10^2 \, \rm kpc$ scales, would hold up to Mpc scale distances. There are currently no available numerical or semi-analytical models which can test its validity of this scenario on the much longer time and spatial scales implied by Porphyrion. At the same time, several more objects with similar characteristics but slightly smaller overall extension were already detected by recent LOFAR surveys \citep[e.g.][]{2023A&A...672A.163O,2024A&A...686A..21S}.

This context makes it interesting to explore our radically different interpretation, which bypasses the anomalies that Porphyrion poses to our understanding of radio jets.
Figure \ref{fig:sketch} gives a schematic view of the most important differences implied by the two models of \por\ discussed in this work.
The \gr\ beam trace model considered above provides such alternative interpretation. Contrary to the conventional jet models, the problem of directional stability does not even exist in the \gr\ beam trace model. \gr s are generically expected to deposit electrons and positrons along a straight line of the length comparable to the \gr\ mean free path. Thus, straight Mpc-scale line-like features like \por\ are good candidates for being traces of \gr\ beams from their parent AGN. 

In principle, straight jet-like features that are electron-positron pair traces of \gr\ beams should be present in all AGNs that emit TeV \gr s. It is also likely that the \gr\ beams in such AGN are ejected in the direction of the black hole spin axis in which case they are expected to be aligned along the directions of the AGN jets.
TeV \gr s can also be produced by high-energy particles accelerated in the jets. A TeV \gr\ beam propagating through the jet may simply provide an additional source of high-energy electrons and positrons in the jet. It is not obvious how to distinguish this source from other possible sources of high-energy particles in the jets such as, e.g., shock acceleration in the jet knots. 

According to the model for Porphyrion, given in \citet{nature}, the jet had an average power of $P \sim 10^{46}\erg\s^{-1}$ active for $\tau \sim 1.9 \rm ~Gyr$, and the total deposited energy is $ \sim 8 \cdot 10^{62}\erg$ in the jets.  This implies a total accreted mass of  $\Delta M \sim 10^9 M_\odot$, which makes the association of \por\  with the quasar in the region (total mass $M_q \sim 2 \cdot 10^{8} M_\odot$) impossible. 

In the \gr\ beam model,  $E\sim 30 \rm ~GeV$ electrons emitting synchrotron radiation  at $100~ \rm MHz$ (the latter being energetically negligible), have been initially injected with $E\sim \rm TeV$ energy by a \gr\ beam. The injection should have lasted at least $\sim 10$~Myr (the time needed for \gr s to propagate over the jet length). The present-day luminosity of \por\ in \gr\ band (Fig.~\ref{fig:sed}) is of $\lesssim 5\times 10^{45}$~erg/s (depending on the unknown magnetic field strength in the jet). This suggests an energy output of $\lesssim 10^{60}$~erg on 10~Myr time scale.  The energy injected by the \gr\ beam must have been  $\sim \rm 1 TeV/30 GeV \sim 33$ times larger in the past, i.e. the total energy of the \gr\ beam should have been $\lesssim 3 \cdot 10^{61} \rm erg$. We can thus estimate an overall $\gtrsim 30$ times lower energy budget to explain \por\ according to the \gr\ beam scenario, compared to the conventional radio jet model. 

Our study of the non-detection of inverse Compton emission from \por\ in \gr s yields a lower bound on the magnetic field in the jet.
This constraint can be further improved via a search of inverse Compton flux at lower energies, as it is clear from Fig.~\ref{fig:sed}. GeV energy electrons are generically expected to be present in the jet because of their relatively long cooling time scale; see Eq.~(\ref{eq:tcool}). Such electrons produce inverse Compton emission in the hard X-ray energy range (\ref{IC}). The $E\sim 100\keV$ inverse Compton emission from such electrons is unfortunately out of the reach of existing and planned telescopes in this energy range.  If the activity time scale of the \gr\ beam from the parent AGN is in the $\gtrsim 100$~Myr range, cooling of electrons to sub-GeV energies may result in the inverse Compton emission in the X-ray range, where the sensitivity of existing telescopes may be sufficient for its detection.

\section{Conclusion}
To summarize, our proposed \gr\ model for Porphyrion presents the following advantages:
(a) it naturally explains the straightness of the jet on such large scales, (b) it makes no critical requirements on the jets' internal physical parameters, (c) it lowers the energy requirements for the central AGN, and (d) it does not require jet propagation in the very low density environment of the LSS voids. 
The model can be observationally tested with future hard X-ray observations, and through the statistical comparison of larger dataset of \por-like systems in future radio surveys.

\section*{Acknowledgments}
We acknowledge fruitful scientific discussion with M.\ Br\"{u}ggen, M.\ Perucho, M.\ Oei, H.\ Rottgering and S. Mtchedlidze. 
F.V.\ acknowledges CINECA, under the award  ``IsB28\_RADGALEO'' under the ISCRA initiative, for the availability of high-performance computing resources and support, and  Fondazione Cariplo and Fondazione CDP, for grant n$^\circ$ Rif: 2022-2088 CUP J33C22004310003 for "BREAKTHRU" project.
A.B.\ acknowledges support by the Swedish Research Council (Vetenskapsr{\aa}det, 2019-04234).

\bibliographystyle{aa}
\bibliography{refs}

\begin{thebibliography}{24}
\expandafter\ifx\csname natexlab\endcsname\relax\def\natexlab#1{#1}\fi

\bibitem[{Acciari {et~al.}(2023)}]{MAGIC:2022piy}
Acciari, V.~A. {et~al.} 2023, \aap, 670, A145

\bibitem[{{Aharonian} {et~al.}(2023){Aharonian}, {Aschersleben}, {Backes},
  {Martins}, {Batzofin}, {Becherini}, {Berge}, {Bi}, {Bouyahiaoui}, {Breuhaus},
  {Brose}, {Brun}, {Bruno}, {Bulik}, {Burger-Scheidlin}, {Bylund}, {Caroff},
  {Casanova}, {Celic}, {Cerruti}, {Chand}, {Chandra}, {Chen}, {Chibueze},
  {Chibueze}, {Cotter}, {de Bony}, {Egberts}, {Ernenwein}, {Fichet de
  Clairfontaine}, {Filipovic}, {Fontaine}, {F{\"u}ssling}, {Funk}, {Gabici},
  {Ghafourizadeh}, {Giavitto}, {Glawion}, {Glicenstein}, {Goswami}, {Grondin},
  {Haerer}, {Holch}, {Holler}, {Horns}, {Jamrozy}, {Jankowsky}, {Joshi},
  {Jung-Richardt}, {Kasai}, {Katarzy{\'n}ski}, {Khatoon}, {Kh{\'e}lifi},
  {Klu{\'z}niak}, {Komin}, {Kosack}, {Kostunin}, {Lang}, {Le Stum}, {Leitl},
  {Lemi{\`e}re}, {Lenain}, {Leuschner}, {Lohse}, {Luashvili}, {Lypova},
  {Mackey}, {Malyshev}, {Malyshev}, {Marandon}, {Marchegiani}, {Marcowith},
  {Mart{\'\i}-Devesa}, {Marx}, {Meyer}, {Mitchell}, {Moderski}, {Mohrmann},
  {Montanari}, {Moulin}, {Muller}, {Murach}, {Nakashima}, {Niemiec}, {Ohm},
  {Olivera-Nieto}, {de Ona Wilhelmi}, {Panny}, {Panter}, {Parsons}, {Peron},
  {Prokhorov}, {Prokoph}, {P{\"u}hlhofer}, {Punch}, {Quirrenbach},
  {Reichherzer}, {Reimer}, {Reimer}, {Reville}, {Rieger}, {Rowell}, {Rudak},
  {Ruiz-Velasco}, {Sahakian}, {Sanchez}, {Sasaki}, {Sch{\"u}ssler}, {Schutte},
  {Schwanke}, {Shapopi}, {Sol}, {Spencer}, {Steinmassl}, {Suzuki}, {Takahashi},
  {Tanaka}, {Taylor}, {Terrier}, {Thorpe-Morgan}, {Tsirou}, {Tsuji},
  {Uchiyama}, {van Eldik}, {Veh}, {Venter}, {Wagner}, {White}, {Wierzcholska},
  {Wong}, {Zacharias}, {Zargaryan}, {Zdziarski}, {Zouari}, {{\.Z}ywucka},
  {Meyer}, \& {Fermi-LAT Collaboration}}]{2023ApJ...950L..16A}
{Aharonian}, F., {Aschersleben}, J., {Backes}, M., {et~al.} 2023, \apjl, 950,
  L16

\bibitem[{Atwood {et~al.}(2009)Atwood, Abdo, Ackermann, Althouse, Anderson,
  Axelsson, Baldini, Ballet, Band, Barbiellini, Bartelt, Bastieri, Baughman,
  Bechtol, Bédérède, Bellardi, Bellazzini, Berenji, Bignami, Bisello,
  Bissaldi, Blandford, Bloom, Bogart, Bonamente, Bonnell, Borgland, Bouvier,
  Bregeon, Brez, Brigida, Bruel, Burnett, Busetto, Caliandro, Cameron, Caraveo,
  Carius, Carlson, Casandjian, Cavazzuti, Ceccanti, Cecchi, Charles, Chekhtman,
  Cheung, Chiang, Chipaux, Cillis, Ciprini, Claus, Cohen-Tanugi, Condamoor,
  Conrad, Corbet, Corucci, Costamante, Cutini, Davis, Decotigny, DeKlotz,
  Dermer, de~Angelis, Digel, do~Couto~e Silva, Drell, Dubois, Dumora, Edmonds,
  Fabiani, Farnier, Favuzzi, Flath, Fleury, Focke, Funk, Fusco, Gargano,
  Gasparrini, Gehrels, Gentit, Germani, Giebels, Giglietto, Giommi, Giordano,
  Glanzman, Godfrey, Grenier, Grondin, Grove, Guillemot, Guiriec, Haller,
  Harding, Hart, Hays, Healey, Hirayama, Hjalmarsdotter, Horn, Hughes,
  Jóhannesson, Johansson, Johnson, Johnson, Johnson, Johnson, Kamae, Katagiri,
  Kataoka, Kavelaars, Kawai, Kelly, Kerr, Klamra, Knödlseder, Kocian, Komin,
  Kuehn, Kuss, Landriu, Latronico, Lee, Lee, Lemoine-Goumard, Lionetto, Longo,
  Loparco, Lott, Lovellette, Lubrano, Madejski, Makeev, Marangelli, Massai,
  Mazziotta, McEnery, Menon, Meurer, Michelson, Minuti, Mirizzi, Mitthumsiri,
  Mizuno, Moiseev, Monte, Monzani, Moretti, Morselli, Moskalenko, Murgia,
  Nakamori, Nishino, Nolan, Norris, Nuss, Ohno, Ohsugi, Omodei, Orlando, Ormes,
  Paccagnella, Paneque, Panetta, Parent, Pearce, Pepe, Perazzo, Pesce-Rollins,
  Picozza, Pieri, Pinchera, Piron, Porter, Poupard, Rainò, Rando, Rapposelli,
  Razzano, Reimer, Reimer, Reposeur, Reyes, Ritz, Rochester, Rodriguez, Romani,
  Roth, Russell, Ryde, Sabatini, Sadrozinski, Sanchez, Sander, Sapozhnikov,
  Parkinson, Scargle, Schalk, Scolieri, Sgrò, Share, Shaw, Shimokawabe,
  Shrader, Sierpowska-Bartosik, Siskind, Smith, Smith, Spandre, Spinelli,
  Starck, Stephens, Strickman, Strong, Suson, Tajima, Takahashi, Takahashi,
  Tanaka, Tenze, Tether, Thayer, Thayer, Thompson, Tibaldo, Tibolla, Torres,
  Tosti, Tramacere, Turri, Usher, Vilchez, Vitale, Wang, Watters, Winer, Wood,
  Ylinen, \& Ziegler}]{Atwood_2009}
Atwood, W.~B., Abdo, A.~A., Ackermann, M., {et~al.} 2009, \apj, 697,
  1071–1102

\bibitem[{Biteau {et~al.}(2020)Biteau, Prandini, Costamante, Lemoine, Padovani,
  Pueschel, Resconi, Tavecchio, Taylor, \& Zech}]{Biteau:2020prb}
Biteau, J., Prandini, E., Costamante, L., {et~al.} 2020, \natas, 4, 124

\bibitem[{{Blandford} {et~al.}(2019){Blandford}, {Meier}, \&
  {Readhead}}]{2019ARA&A..57..467B}
{Blandford}, R., {Meier}, D., \& {Readhead}, A. 2019, \araa, 57, 467

\bibitem[{{Bondarenko} {et~al.}(2022){Bondarenko}, {Boyarsky}, {Korochkin},
  {Neronov}, {Semikoz}, \& {Sokolenko}}]{bond22}
{Bondarenko}, K., {Boyarsky}, A., {Korochkin}, A., {et~al.} 2022, \aap, 660,
  A80

\bibitem[{Carretti {et~al.}(2022)Carretti, O\textquoteright{}Sullivan, Vacca,
  Vazza, Gheller, Vernstrom, \& Bonafede}]{Carretti:2022fqk}
Carretti, E., O\textquoteright{}Sullivan, S.~P., Vacca, V., {et~al.} 2022,
  \mnras, 518, 2273

\bibitem[{Colberg {et~al.}(2005)Colberg, Krughoff, \& Connolly}]{filaments}
Colberg, J.~M., Krughoff, K.~S., \& Connolly, A.~J. 2005, \mnras, 359, 272

\bibitem[{Durrer \& Neronov(2013)}]{Durrer:2013pga}
Durrer, R. \& Neronov, A. 2013, \aar, 21, 62

\bibitem[{Fitoussi {et~al.}(2017)Fitoussi, Belmont, Malzac, Marcowith,
  Cohen-Tanugi, \& Jean}]{Fitoussi_2017}
Fitoussi, T., Belmont, R., Malzac, J., {et~al.} 2017, \mnras, 466, 3472–3487

\bibitem[{Gould \& Schr\'eder(1967)}]{PhysRev.155.1404}
Gould, R.~J. \& Schr\'eder, G.~P. 1967, \pr, 155, 1404

\bibitem[{{Komissarov} \& {Porth}(2021)}]{2021NewAR..9201610K}
{Komissarov}, S. \& {Porth}, O. 2021, \nar, 92, 101610

\bibitem[{{Mtchedlidze} {et~al.}(2024){Mtchedlidze},
  {Dom{\'\i}nguez-Fern{\'a}ndez}, {Du}, {Carretti}, {Vazza}, {O'Sullivan},
  {Brandenburg}, \& {Kahniashvili}}]{2024arXiv240616230M}
{Mtchedlidze}, S., {Dom{\'\i}nguez-Fern{\'a}ndez}, P., {Du}, X., {et~al.} 2024,
  ApJ, in press, arXiv:2406.16230

\bibitem[{Neronov {et~al.}(2002)Neronov, Semikoz, Aharonian, \&
  Kalashev}]{Neronov_2002}
Neronov, A., Semikoz, D., Aharonian, F., \& Kalashev, O. 2002, \prl, 89

\bibitem[{{Neronov} \& {Semikoz}(2009)}]{2009PhRvD..80l3012N}
{Neronov}, A. \& {Semikoz}, D.~V. 2009, \prd, 80, 123012

\bibitem[{{Neronov} \& {Vovk}(2010)}]{2010Sci...328...73N}
{Neronov}, A. \& {Vovk}, I. 2010, \sci, 328, 73

\bibitem[{{Oei} {et~al.}(2024){Oei}, {Hardcastle}, {Timmerman}, {Gast},
  {Botteon}, {Rodriguez}, {Stern}, {Van Weeren}, {Rottgering}, {Stern}, \&
  {Djorgovski}}]{nature}
{Oei}, M., {Hardcastle}, M., {Timmerman}, R., {et~al.} 2024, \nat, 633, 537

\bibitem[{{Oei} {et~al.}(2023){Oei}, {van Weeren}, {Gast}, {Botteon},
  {Hardcastle}, {Dabhade}, {Shimwell}, {R{\"o}ttgering}, \&
  {Drabent}}]{2023A&A...672A.163O}
{Oei}, M. S.~S.~L., {van Weeren}, R.~J., {Gast}, A. R.~D.~J.~G.~I.~B., {et~al.}
  2023, \aap, 672, A163

\bibitem[{{Perucho} \& {L{\'o}pez-Miralles}(2023)}]{2023JPlPh..89e9101P}
{Perucho}, M. \& {L{\'o}pez-Miralles}, J. 2023, \jpp, 89, 915890501

\bibitem[{{Perucho} {et~al.}(2022){Perucho}, {Mart{\'\i}}, \&
  {Quilis}}]{2022MNRAS.510.2084P}
{Perucho}, M., {Mart{\'\i}}, J.-M., \& {Quilis}, V. 2022, \mnras, 510, 2084

\bibitem[{{Simonte} {et~al.}(2024){Simonte}, {Andernach}, {Br{\"u}ggen},
  {Miley}, \& {Barthel}}]{2024A&A...686A..21S}
{Simonte}, M., {Andernach}, H., {Br{\"u}ggen}, M., {Miley}, G.~K., \&
  {Barthel}, P. 2024, \aap, 686, A21

\bibitem[{{Tjemsland} {et~al.}(2024){Tjemsland}, {Meyer}, \& {Vazza}}]{tj24}
{Tjemsland}, J., {Meyer}, M., \& {Vazza}, F. 2024, \apj, 963, 135

\bibitem[{{Vazza} {et~al.}(2019){Vazza}, {Ettori}, {Roncarelli}, {Angelinelli},
  {Br{\"u}ggen}, \& {Gheller}}]{va19}
{Vazza}, F., {Ettori}, S., {Roncarelli}, M., {et~al.} 2019, \aap, 627, A5

\bibitem[{{Vernstrom} {et~al.}(2023){Vernstrom}, {West}, {Vazza}, {Wittor},
  {Riseley}, \& {Heald}}]{2023SciA....9E7233V}
{Vernstrom}, T., {West}, J., {Vazza}, F., {et~al.} 2023, \scia, 9, eade7233

\end{thebibliography}

\end{document}